# A quantized physical framework for understanding the working mechanism of ion channels

Hao Wang[1*], Jiahui Wang[1], Xin Yuan Thow[2], Chengkuo Lee[1*]

[1]Department of Electrical and Computer Engineering, National University of Singapore, Singapore 117583

[2]Singapore Institute for Neurotechnology (SINAPSE), National University of Singapore, Singapore 117456

Correspondence:

Chengkuo Lee

elelc@nus.edu.sg

**Abstract**

A quantized physical framework, called the five-anchor model, is developed for a general understanding of the working mechanism of ion channels. According to the hypotheses of this model, the following two basic physical principles are assigned to each anchor: the polarity change induced by an electron transition and the mutual repulsion and attraction induced by an electrostatic force. Consequently, many unique phenomena, such as fast and slow inactivation, the stochastic gating pattern and constant conductance of a single ion channel, the difference between electrical and optical stimulation (optogenetics), nerve conduction block and the generation of an action potential, become intrinsic features of this physical model. Moreover, this model also provides a foundation for the probability equation used to calculate the results of electrical stimulation in our previous C-P theory.

**Introduction**

Since the discovery of ion channels in 1952 [1], ion channel research has always been the most fundamental topic in neuroscience. Starting from the famous Hodgkin-Huxley model (H-H model), people have tried to build many phenomenological and mathematical models to explain and study the unique properties of all kinds of ion channels [2-17]. Meanwhile, various kinds of experiments and decades of investigation revealed some features, such as the fast inactivation, which is linked to the ball-and-chain domain [18-31], and the stochastic gating pattern of a single ion channel [3,17,32-51]. However, none of the previous models were built from first principles, preventing a full understanding of the physical mechanisms underlying these phenomena. On the other hand, the fragmented experimental observations provide only a glimpse of the complete theory but do allow us to physically rebuild the theory. How can we obtain the complete physical model of ion channels?

An ion channel is a system consisting of several components, and the working process is dominated by some specific physical principles. All the unique phenomena, such as fast and slow inactivation, ion selectivity and the stochastic gating pattern, are just results observed during the working process of this system. By constructing a suitable system with proper and reasonable physical principles, we can not only derive the observed phenomena theoretically but also predict unknown properties of ion channels.

Therefore, in this study, a quantized physical system, called the five-anchor model, is built based on previous molecular studies of ion channel structures [20,28,31,52-79]. In this system, two basic physical principles are assigned to each anchor, the polarity change induced by an electron transition and the mutual repulsion and attraction induced by an electrostatic force. Consequently, many unique phenomena, such as fast and slow inactivation, the stochastic gating pattern and constant conductance of a single ion channel, the difference between electrical and optical stimulation (optogenetics), nerve conduction block and the generation of an action potential, become intrinsic features of this physical model. Moreover, an equation for calculating the activation probability of a single ion channel under a specific electric input can be directly derived and follows the same format of the equation in our previous circuit-probability theory (C-P theory) [80]. Thus, the probability calculus equation, which was merely a hypothesis in the previous theory, can now be derived from a more fundamental physical model.

**Five-anchor model**

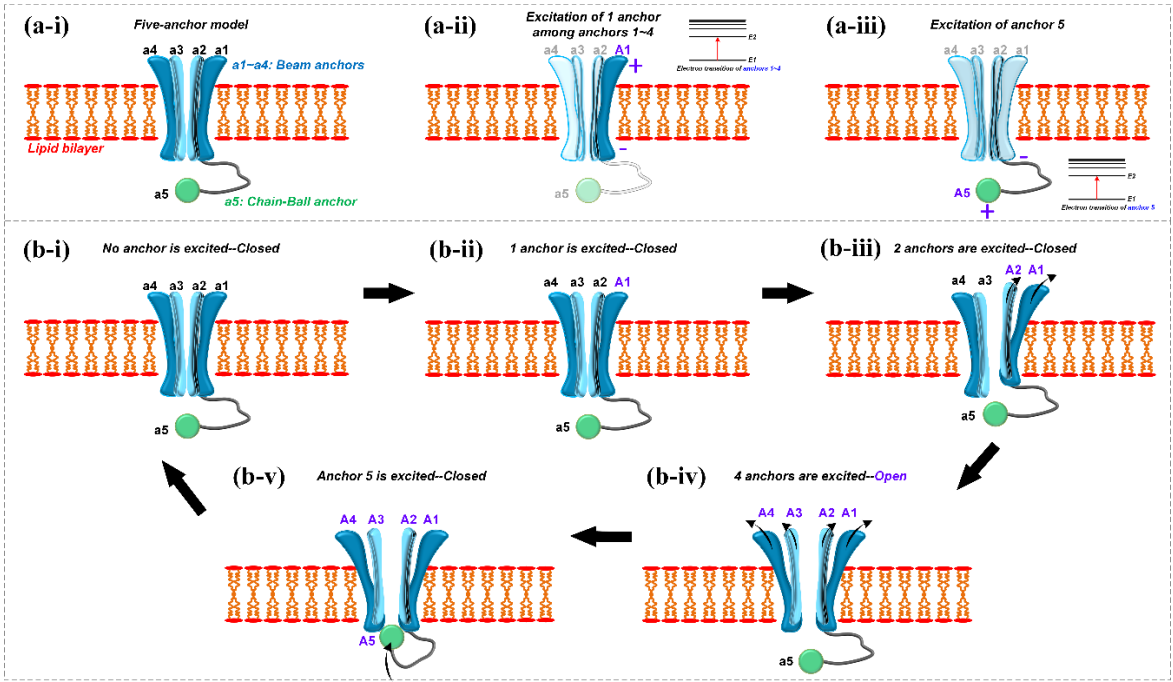

Figure 1. Five-anchor model of an ion channel. (a) Structure of an ion channel and excitation states of the beam anchors and the ball-and-chain anchor: (i) a1 to a4 refer to the four beam anchors, a5 refers to the ball-and-chain anchor, and lowercase indicates the unexcited state; (ii) 1 of the beam anchors is excited, its polarity is changed to a positive top end with a negative bottom end, and uppercase indicates the excited state; (iii) the ball-and-chain anchor is excited, its polarity is changed, the ball becomes positively charged, the chain becomes negatively charged, and uppercase indicates the excited state. (b) A common working process for ion channel gating: (i) no anchor is excited, and the ion channel is closed; (ii) 1 beam anchor is excited, and the ion channel is closed; (iii) 2 beam anchors are excited and repel each other, making these two anchors separate and rise while the channel remains closed; (iv) all four beam anchors are excited and repel each other, causing all of these anchors to separate and rise. The ion channel is open. (v) The ball-and-chain anchor is excited. The positively charged ball will be attracted to the negatively charged bottom end of the beam anchors.

### Physical working process of an ion channel

The five-anchor model is built based on the observed molecular structures of sodium and potassium ion channels and describes a basic physical working process for an individual ion channel. Then, based on the physical working process, a mathematical description of the transition between the closed and open states can be directly obtained.

A typical biological structure of an individual ion channel is shown in Figure 1(a-i), with four protein beams and a ball-and-chain domain. All four protein beams and the ball-and-chain domain are simplified as anchors that control the gating of the ion channel. The four protein beams are named beam anchors (a1 to a4), while the ball-and-chain domain is named the ball-and-chain anchor (a5). An essential point to emphasize here is that we do not specify the exact type of ion channel since this basic model provides only a physical framework to understand of any kind of ion channel that shares a similar structure. The only differences among various kinds of ion channels are the number of protein beams and the presence or absence of the ball-and-chain domain.

In response to external excitation, these anchors can be excited by an electron transition. Here, we assume that an electron in one anchor can be excited from a low energy level to a high energy level (shown in Figure 1(a-ii & a-iii)), resulting in a polarity change in the anchor. For a1 to a4, the excitation will make the top end positively charged and the bottom end negatively charged, as shown in Figure 1(a-ii). For a5, the excitation will make the ball positively charged and the chain negatively charged, as shown in Figure 1(a-iii). Here, the anchors with unexcited and excited states are indicated in lowercase and uppercase, respectively. The essential point to emphasize here is that the excitation of each anchor is independent of the excitation of the other anchors. Thus, the change in the state of one anchor (for example, from excited to unexcited or from unexcited to excited) is determined only by its current state and the external excitation and is not affected by the states of the other anchors.

A typical gating process for action potential generation is shown in Figure 1(b). In the initial step, all 5 anchors are unexcited; thus, the ion channel is closed (Figure 1(b-i)). The first beam anchor is then excited and changes polarity, but nothing happens, and the ion channel is still closed (Figure 1(b-ii)). Next, a second anchor is excited and changes polarity. Both excited anchors are of the same polarity, causing repulsion between the anchors and making these two anchors separate and rise (Figure 1(b-ii)). However, in this step, the ion channel is still closed. The ion channel opens only when all four beam anchors are excited and repel each other (Figure 1(b-iv)). The ball-and-chain anchor is then excited. Since the ball is positively charged, it will be attracted to the negatively charged bottom end of the excited beam anchors and block the ion channel (Figure 1(b-v)). In this step, the ion channel is closed. After a certain duration, all anchors return to the unexcited state (Figure 1(b-i)), and the ion channel is closed.

## Mathematical description

Since the anchor excitation is induced by an electron transition, a mathematical description of the state transition can also be directly obtained from this physical mechanism.

## Excitation by an electric field

For an electron transition induced by an electric field, there will also be a threshold strength. Thus, as observed experimentally, there will be a threshold voltage for exciting the anchors [16].

Here, define the event of excitation of the anchor $n$ as $A_n$ and the recovery of the anchor $n$ as $a_n$. Considering all quantum events follow the exponential distribution, then the probability of $A_n$ is:

$$P(A_n) = 1 - e^{-\int \lambda_{An}(V(t))dt};$$

Here, the exponential distribution has been rewritten in its integral format.

The function $\lambda_{An}$ is:

$$\lambda_{An}(V(t)) = \alpha_n \times \frac{1}{e^{\frac{\beta_n}{|V(t) - V_{T_n}|}} - 1};$$

Here, $V_{T_n}$ refers to the threshold voltage of the anchor $n$. $\alpha_n$ and $\beta_n$ are physical parameters determined by the molecular structure of the anchor $n$. $V(t)$ refers to the effective voltage on the ion channel, which can be introduced by any source, such as an action potential, electrical stimulation, magnetic stimulation and even acoustic stimulation. The process by which magnetic stimulation and acoustic stimulation can apply voltage to an ion channel is explained in our previous study [81]. This equation follows the same form as Planck's law since the electron transition is a quantum event.

The probability of $a_n$, the recovery event for anchor $n$, is:

$$P(a_n) = 1 - e^{-\int \lambda_{an} dt} = 1 - e^{-\lambda_{an} \times t};$$

Here, $\lambda_{an}$ is a constant and not affected by the external voltage. Thus, the transition of the electron from a higher energy to a lower energy, which is a spontaneous process, cannot be controlled by an external input. Quantum mechanically, this recovery still can be controlled by another photon, called stimulated emission [82], but this mechanism is neglected here. Thus,

$$\lambda_{an} \propto \alpha_n;$$

The physical meaning of this formula is that if the electron transition is more likely to happen (the energy level difference is lower), the time for the electron to return from the high energy level to the low energy level will be shorter.

This effect on the ball-and-chain anchor (a5) will be quite particular. Once a5 is excited and blocks the ion channel, this ion channel cannot be opened by external stimulation again until a5 returns to the unexcited state. **This phenomenon is observed as the refractory period [1] and is also called fast inactivation.**

Since all parameters involved are determined by the molecular structure, we can assume that the beam anchors (a1~a4) share the same parameters, which are different from those of ball-and-chain anchor (a5). Moreover, the function of the ball-and-chain anchor is to close the ion channel when the amplitude of the action potential exceeds a specific level, so a5 should be triggered at a relatively higher voltage. Thus, the threshold voltage of a5 ($V_{T_5}$) should be higher than that of beam anchors.

With the above equations, the gating process of a single ion channel under an exact external applied voltage waveform can be modeled as follow.

The opening of an ion channel (event $O$) can be expressed as:

$$O = A_1 \cap A_2 \cap A_3 \cap A_4 \cap a_5;$$

(All beam anchors are excited & the ball-and-chain anchor is unexcited)

The closing of an ion channel (event $C$) is:

$$C = a_1 \cup a_2 \cup a_3 \cup a_4 \cup A_5;$$

(At least one beam anchor is unexcited or the ball-and-chain anchor is excited)

Electrical stimulation becomes a special case, which is the state transition from the initial state ($a_1 \cap a_2 \cap a_3 \cap a_4 \cap a_5$) to the open state ($A_1 \cap A_2 \cap A_3 \cap A_4 \cap a_5$). The equation to calculate the probability of excitation is:

$$P(A_1) \times P(A_2) \times P(A_3) \times P(A_4) = \left(1 - e^{-\int \lambda_{An}(V(t))dt}\right)^4 = (1 - e^{-\int \alpha_n \times \frac{1}{e^{\frac{\beta_n}{|V(t) - V_{T_n}|}}-1}dt})^4;$$

This equation follows the same format used in our previous C-P theory for calculating the result of electrical nerve stimulation. Therefore, the probability calculation equation can now be derived from a more fundamental physical model.

Previously, people already observed the stochastic gating pattern of a single ion channel [32-51]. Here, we can reproduce this gating pattern, as shown in Figure 2.

As shown, with increases in the DC voltage (from V=-1 to V=-7), the ion channel will open more frequently. However, the ion channel cannot always remain in the open state since the beam anchors can spontaneously return to an unexcited state. However, with a higher voltage, the closed duration will be shorter because the deexcited beam anchor can be quickly excited again and make the ion channel open.

Additionally, there will be some very short open periods. This situation occurs when one beam anchor returns to an unexcited state immediately after the excitation of another deexcited beam. This very short open period can always happen and is not controllable [32-51].

Additionally, in the measurement of single ion channel gating, an interesting result is that the conductance of a single ion channel remains constant while the applied voltage increases [17,35-36,43,46-48]. This phenomenon can now be easily understood in the new model since the voltage, or electric field, is not the direct factor that opens the ion channel. The electric field triggers the electron transition only. Consequently, the conductance of a single ion channel, which is determined by the pore dimensions, is determined by the mutual repulsion, which is independent of the external applied electric field. Thus, the conductance of a single ion channel should be a constant value that is not affected by the applied voltage.

Upon further increase in the voltage, the ball-and-chain anchor is excited, closing the ion channel. Thus, as seen, the channel is now closed more frequently. However, regardless of the DC voltage level, there will always be some short open periods since a5 will spontaneously recover and enable opening of the ion channel. However, in our theory, the voltage $V(t)$ is not necessarily a DC voltage. Actually, the gating pattern of any voltage can be modeled. If a high frequency AC voltage with different amplitudes is applied, the gating pattern will be similar to the pattern for a DC voltage, as shown in Figure 2.

Thus, with the present model, we can directly obtain a prediction for nerve conduction block. Nerve conduction block is a phenomenon observed when a continuous electrical input is applied to a nerve [71,83-103]. This continuous electrical input can block the neural signal propagation.

It is easier to understand this phenomenon with Figure 2. When a voltage is high enough to keep exciting the beam anchors, which is the case when V=-7, the ion channel is always activated, resulting in two effects. First, the ion channel cannot respond to an incoming action potential since all beam anchors remain triggered. Additionally, the frequent opening of the ion channel will deplete the ion concentration difference, which is essential for the activation of action potential. Thus, for this case of nerve conduction block, there will be substantial fatigue after the blocking test. However, if a further increase in the voltage is applied in the blocking test, the gating of the ion channel will change from frequently open to frequently closed because of the excitation of a5. Therefore, in this case, the ion channel also cannot respond to an incoming action potential. However, since the ion channel is always closed, there will be no depletion of the ion gradient and thus no fatigue after the blocking test. The resulting transition from serious fatigue to no fatigue in response to an increase in the voltage amplitude has been observed and confirmed in a previous study [94].

Moreover, it is easy to predict that the voltage or current required for DC nerve conduction block will be much higher than that for AC nerve conduction block. As explained in our previous study [81], the neural circuit has a parallel RLC circuit configuration. Thus, its voltage response to an AC input will be much higher than the response to a DC input. Consequently, a higher DC input is required to reach the threshold voltage for triggering the anchors.

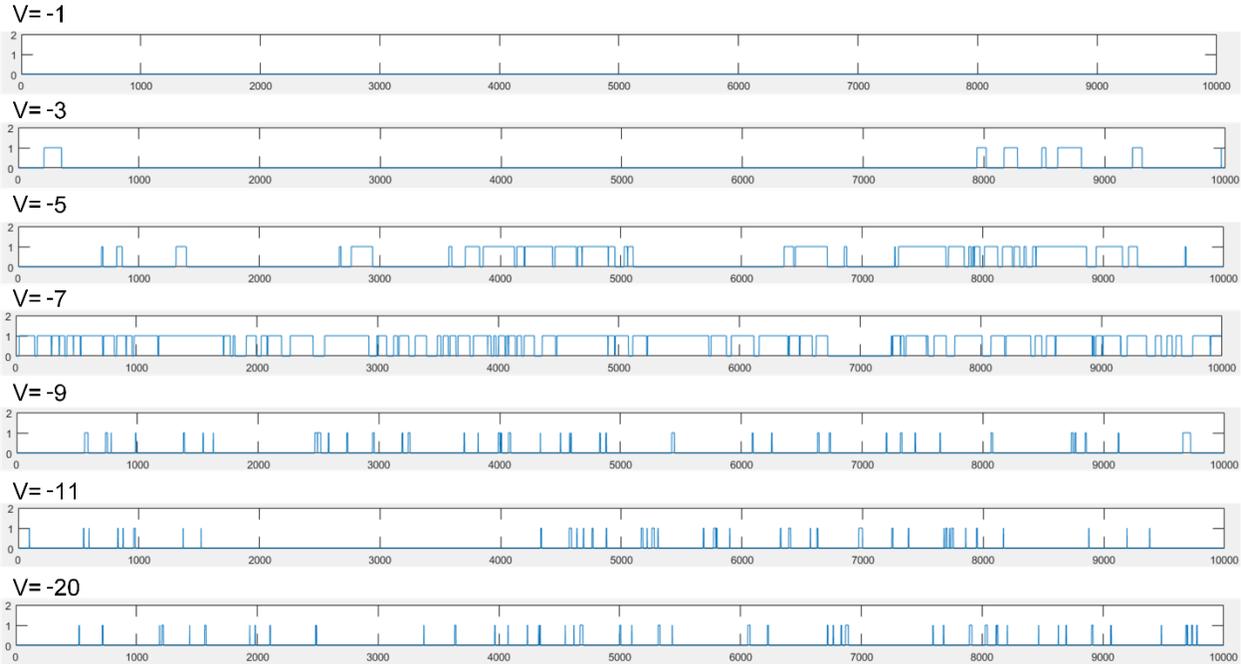

Figure 2. The gating pattern of a single ion channel with different DC voltages.

**Excitation by a photon**

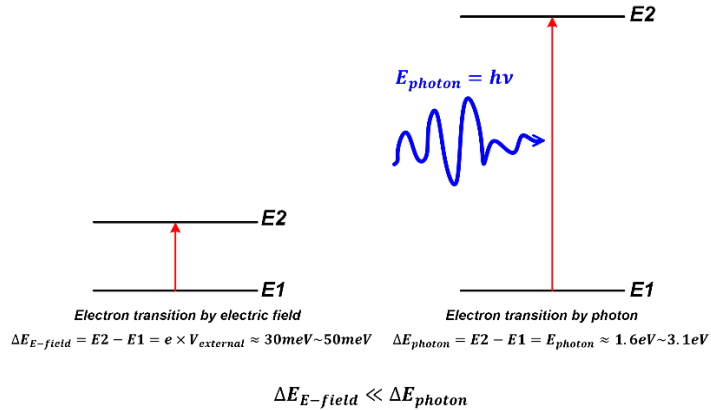

Figure 3 Comparison of the electron transition in conditions of electrical nerve stimulation and optical nerve stimulation.

The electron transition induced by a photon is the well-known photoelectric effect. To induce an electron transition from a low energy level to a high energy level, the energy of the photon should be higher than the energy level difference. Therefore, only light whose frequency is higher than the threshold can excite the electron transition and thus trigger the opening an ion channel. This process is what happens in optogenetic neural stimulation [104-113]. Although voltage-gated ion channels do not have exactly the same structure as light-gated ion channels, the fundamental working mechanism, which is the electron transition, is considered to be the same. Thus, similar to the criteria for the photoelectric effect, the frequency of the light should be higher than the threshold to activate light-gated ion channels [106].

Circumstantial evidence supporting this opinion comes from the excited state lifetime difference between electrical nerve stimulation and optical nerve stimulation, as explained in Figure 3.

For electrical nerve stimulation, the typical external voltage required ranges from 30 mV to 50 mV. The energy absorbed by the electron to transition from a lower energy level to a higher energy level can be estimated as:

$$\Delta E_{E-field} = E2 - E1 = e \times V_{external} \approx 30\ meV \sim 50\ meV$$

However, the light required for optogenetic nerve stimulation is normally within the visible light range, with wavelengths from 400 nm to 760 nm. Thus, the energy delivered to the electron can be estimated as:

$$\Delta E_{photon} = E2 - E1 = E_{photon} = h\nu \approx 1.6\ eV \sim 3.1\ eV$$

As seen, compared with the energy for electrical stimulation, the energy required for optical stimulation is much higher, resulting in a much longer excited state lifetime. Thus, the electron can remain in the high energy level for a longer time before spontaneous emission happens. Therefore, the time for each anchor to return from the excited state is longer. Before the recovery of all anchors, the ion channel cannot be stimulated again. Thus, as observed experimentally, optogenetic neural stimulation normally has a very long "off kinetics" time, ranging from 4 ms to 29 min, while an ion channel stimulated by an electric field can normally recover within 1 ms [104,106].

**Explanation for the generation of an action potential with the five-anchor model**

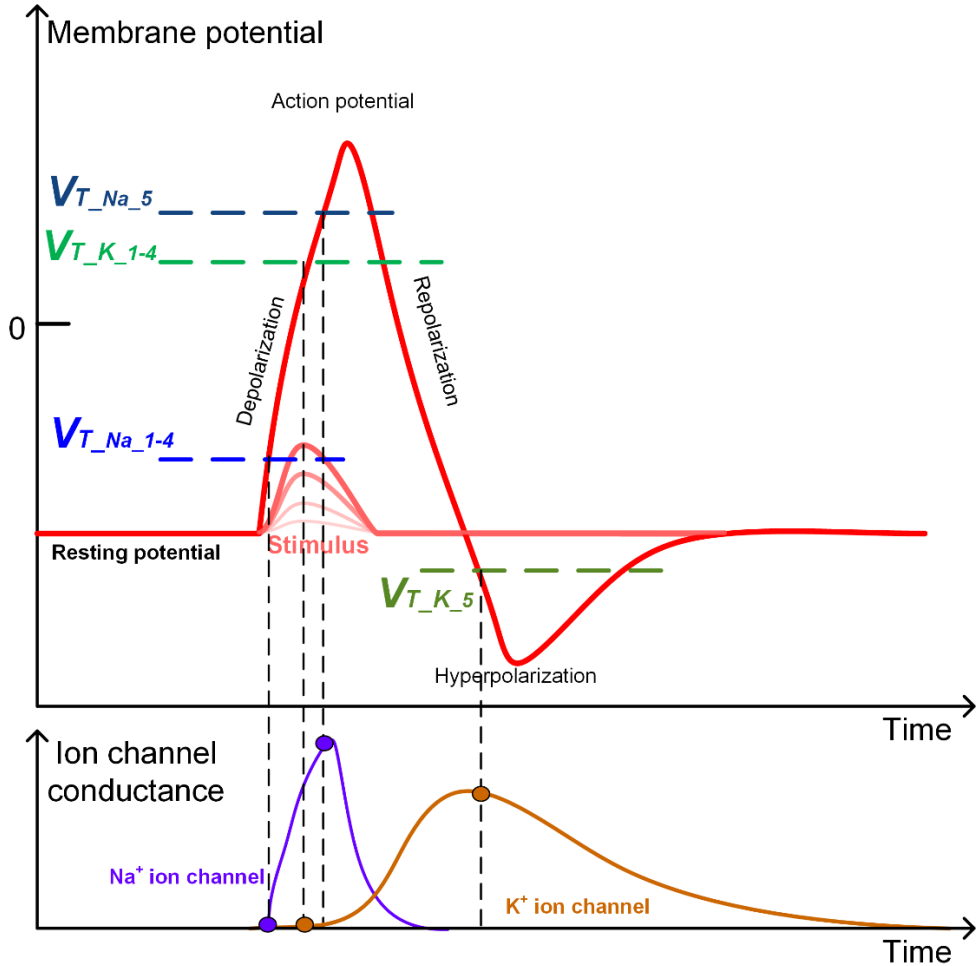

Figure 4 Explanation for the generation of an action potential with the five-anchor model. $V_{T\_Na\_1-4}$ is the threshold voltage of the four beam anchors of a Na ion channel. $V_{T\_Na\_5}$ is the threshold voltage of the ball-and-chain anchor of a Na ion channel. $V_{T\_K\_1-4}$ is the threshold voltage of the four beam anchors of a K ion channel. $V_{T\_K\_5}$ is the threshold voltage of the ball-and-chain anchor of a K ion channel.

To explore the mechanism underlying the operation of the nervous system, many biological neuron models have been developed [1,114-135]. Most of these models provide analytical expressions, enabling people to adjust the parameters for various types of nerves and situations to fit the waveform of the action potential and its firing rate. However, for the five-anchor model, the action potential cannot be described analytically but can be described only numerically due to the involvement of multiple physical processes. The details are explained below.

In the five-anchor model, the generation of an action potential can be understood as shown in Figure 4:

1.  For a stimulus that is higher than the threshold voltage of the four beam anchors of a Na ion channel, $V_{T\_Na\_1-4}$, there is some probability that a Na ion channel will open. The opening of one Na ion channel will induce a Na$^+$ ion flow to further increase the potential and open more Na ion channels. The increasing potential is called depolarization. In this stage, the measured conductance of the Na ion channels will increase because of the increasing number of open Na ion channels.

2. If the increasing membrane potential reaches the threshold voltage for the four beam anchors of a K ion channel, $V_{T\_K\_1-4}$, there is some probability that a K ion channel will open. Therefore, the conductance of the K ion channel increases. However, due to the high conductance of the Na ion channels, the Na$^+$ ion flow is much higher than the K$^+$ ion flow; thus, the membrane potential keeps increasing.

3. The membrane potential reaches the threshold voltage for the ball-and-chain anchor of a Na ion channel, $V_{T\_Na\_5}$. All the Na ion channels will gradually close. The measured conductance of the Na ion channels will decrease.

4. The dominant ion flow changes from Na$^+$ to K$^+$, and the membrane potential reaches the peak value and then decreases. This voltage decrease is called repolarization.

5. The decreasing voltage reaches the threshold voltage of the ball-and-chain anchor of a K ion channel, $V_{T\_K\_5}$. All the K ion channels will gradually close. The measured conductance of the K ion channels will decrease.

6. The membrane potential reaches the lowest value and then recovers via the leakage K ion channel. This process is called hyperpolarization.

As seen, the generation of a complete action potential consists of several different physical processes. Although a mathematical description exists for each individual process, the whole process cannot be expressed analytically. However, even without a complete mathematical equation, using just the five-anchor model, we still can understand why the action potential should generally follow this shape. However, due to the stochastic nature of a single ion channel, the whole process is probabilistic rather than deterministic, which means that the shape of the action potential generated by a single axon is not always identical. Thus, an accurate mathematical equation to predict the shape of an action potential is actually unavailable.

**The understanding of slow activation**

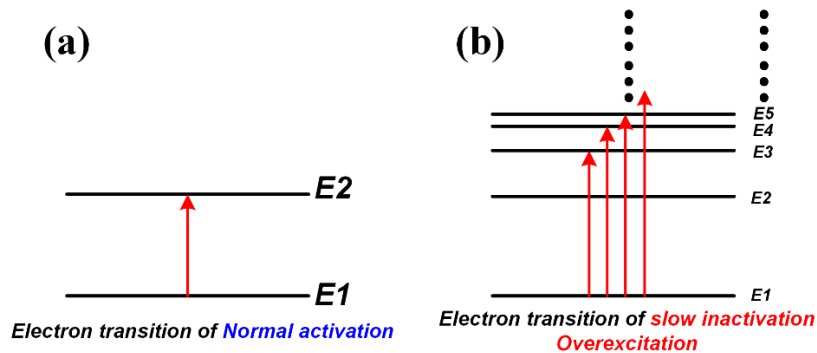

Figure 5. Electron transition of the beam anchors for (a) normal activation and (b) slow inactivation.

Slow inactivation of voltage-gated Na$^+$ channels is a gating phenomenon that is distinct from fast inactivation; with slow inactivation, prolonged depolarizing pulses reduce the number of channels available to provide an excitatory inward current [136]. A hallmark of slow inactivation is that its onset and recovery take place over time domains that span multiple orders of magnitude, such that channels subjected to longer depolarizing pulses require longer times at a negative potential for the current to be restored [137-140].

This phenomenon can also be easily understood with the five-anchor model. As explained above and shown in Figure 5(a), activation of the beam anchor is induced by the electron transition from E1 to E2. However, multiple energy levels are higher than E1. Thus, if a too high voltage is applied to the ion channel, it is possible to make the electron transition from E1 to an energy level that is higher than E2, as shown in Figure 5(b), and this condition is called overexcitation.

Currently, we cannot know the exact consequences of this overexcitation since the atomic orbital is very complicated. However, we still can make some conjectures. In this five-anchor model, the definition of anchor excitation is that an anchor reaches the proper polarization condition after the electron transition. The proper working mechanism completely relies on this proper polarization. If overexcitation happens, this proper polarization can also be jeopardized. As a result, the polarity may be flipped; thus, the beam anchors would attract each other rather than repel each other. An extreme situation is that the electron is excited to the vacuum level (the molecule is ionized to lose this electron); then, the molecule would be charged rather than polarized, also preventing the mutual repulsion between beam anchors.

Consequently, it is easy to understand why this slow inactivation happens when a too long depolarization pulse is applied [137-140]. This situation occurs because a prolonged depolarizing duration will make the voltage too high and enable overexcitation. As long as the electron is excited to an energy level other than E2, slow inactivation will occur. Since there are so many energy levels higher than E2, slow inactivation is a group of events including many possibilities rather than a single event. As we mentioned above, the excited state lifetime is determined by the energy level difference. A higher energy level difference will induce a longer excited state lifetime. Thus, with the increase in the energy level difference, the recovery time can span multiple orders of magnitude, which is very similar as the "off kinetics" of optogenetics.

Fast inactivation has been observed to be independent of slow activation. Removal of fast inactivation not only left slow inactivation intact but also apparently facilitated the process [140]. This phenomenon can also be easily understood. As explained before, fast inactivation is induced by the excitation of anchor 5, which is the ball-and-chain domain. The function of anchor 5 is to maintain the amplitude of the action potential within a certain range. When anchor 5 is removed, there is no control of the amplitude of the action potential. Consequently, the action potential can become higher than the threshold voltage and can induce overexcitation.

## Summary

In summary, a quantized physical model, called the five-anchor model, is proposed to offer a framework for understanding the physical working processes of an individual ion channel. According to the hypotheses of this model, the following two basic physical principles are assigned to each anchor: a polarity change induced by an electron transition and a mutual repulsion and attraction induced by an electrostatic force. Many unique phenomena, such as fast and slow inactivation, the stochastic gating pattern and constant conductance of a single ion channel, the difference between electrical and optical stimulation (optogenetics), nerve conduction block and the generation of an action potential, become intrinsic features of this physical model. Furthermore, this model also provides a foundation for the probability equation used to calculate the results of electrical stimulation in our previous C-P theory.

## Acknowledgement

This work was supported by grants from the National Research Foundation Competitive research programme (NRF-CRP) 'Peripheral Nerve Prostheses: A Paradigm Shift in Restoring Dexterous Limb Function' (NRF-CRP10-2012-01), National Research Foundation Competitive research programme (NRF-CRP) 'Energy Harvesting Solutions for Biosensors' (R-263-000-A27-281), National Research Foundation Competitive research programme (NRF-CRP) 'Piezoelectric Photonics Using CMOS Compatible AlN

Technology for Enabling The Next Generation Photonics ICs and Nanosensors' (R-263-000-C24-281), Faculty Research Committee (FRC) 'Thermoelectric Power Generator (TEG) Based Self-Powered ECG Plaster - System Integration (Part 3)' (R-263-000-B56-112) and HIFES Seed Funding 'Hybrid Integration of Flexible Power Source and Pressure Sensors' (R-263-501-012-133).

**Author contribution**

The theory was developed by Hao Wang. The modeling work was carried out by Hao Wang and Jiahui Wang. The manuscript was written by Hao Wang. All authors contributed to the final version of the manuscript.

Prof. Chengkuo Lee provided general guidance and supervision of the project.

**Conflict of Interest Statement**

The authors certify that they have NO affiliations with or involvement in any organization or entity with any financial interest or non-financial interest in the subject matter or materials discussed in this manuscript.